\documentclass{pasj02}
\Received{$\langle$reception date$\rangle$}
\Accepted{$\langle$acception date$\rangle$}
\Published{$\langle$publication date$\rangle$}

\usepackage{natbib}
\usepackage{amsmath}
\usepackage[switch,mathlines]{lineno}
\usepackage{url}

\begin{document}

\title{XRISM observations of solar flare X-ray emission reflected in the Earth's atmosphere}

\author{Hiromasa Suzuki\altaffilmark{1, 2}\altemailmark}
\author{Jun Kurashima\altaffilmark{2}\altemailmark}
\author{Satoru Katsuda\altaffilmark{3}\altemailmark}
\author{Koji Mori\altaffilmark{2}\altemailmark}
\author{Shun Inoue\altaffilmark{4}}
\author{Daiki Ishi\altaffilmark{1}}
\author{Eugene M. Churazov\altaffilmark{5, 6}}
\author{Ildar Khabibullin\altaffilmark{5, 6, 7}}
\author{Rashid A. Sunyaev\altaffilmark{5, 6}}
\author{Tsunefumi Mizuno\altaffilmark{8}}
\author{Caroline Kilbourne\altaffilmark{9}}
\author{Yuichiro Ezoe\altaffilmark{10}}
\author{Hiroshi Nakajima\altaffilmark{11}}
\author{Kosuke Sato\altaffilmark{12, 13}}
\author{Eric Miller\altaffilmark{14}}
\author{Kyoko Matsushita\altaffilmark{15}}

\altaffiltext{1}{Institute of Space and Astronautical Science (ISAS), Japan Aerospace Exploration Agency (JAXA), Kanagawa 252-5210, Japan} 
\altaffiltext{2}{Faculty of Engineering, University of Miyazaki, Miyazaki 889-2192, Japan} 
\altaffiltext{3}{Department of Physics, Saitama University, Saitama 338-8570, Japan} 
\altaffiltext{4}{Department of Physics, Graduate School of Science, Kyoto University, Kitashirakawa Oiwake-cho, Sakyo-ku, Kyoto 606-8502, Japan}
\altaffiltext{5}{Max Planck Institute for Astrophysics, Karl-Schwarzschild-Str. 1, 85741 Garching, Germany}
\altaffiltext{6}{Space Research Institute (IKI), Profsoyuznaya 84/32, Moscow 117997, Russia}
\altaffiltext{7}{Universitäts-Sternwarte, Fakultät für Physik, Ludwig-Maximilians-Universität München, Scheinerstr.1, 81679 München, Germany}
\altaffiltext{8}{Hiroshima Astrophysical Science Center, Hiroshima University, Hiroshima 739-8526, Japan} 
\altaffiltext{9}{NASA / Goddard Space Flight Center, Greenbelt, MD 20771, USA} 
\altaffiltext{10}{Department of Physics, Tokyo Metropolitan University, Tokyo 192-0397, Japan} 
\altaffiltext{11}{College of Science and Engineering, Kanto Gakuin University, Kanagawa 236-8501, Japan} 
\altaffiltext{12}{International Center for Quantum-field Measurement Systems for Studies of the Universe and Particles (QUP), KEK, Tsukuba 305-0801, Japan}
\altaffiltext{13}{Department of Astrophysics and Atmospheric Sciences,
Kyoto Sangyo University, Kamigamo-motoyama, Kita-ku, Kyoto 603-8555, Japan}
\altaffiltext{14}{Kavli Institute for Astrophysics and Space Research, Massachusetts Institute of Technology, MA 02139, USA} 
\altaffiltext{15}{Faculty of Physics, Tokyo University of Science, Tokyo 162-8601, Japan} 


\email{hiromasa050701@gmail.com, jun@astro.miyazaki-u.ac.jp, katsuda@mail.saitama-u.ac.jp, mori@astro.miyazaki-u.ac.jp}


\maketitle

\begin{abstract}
The X-ray Imaging and Spectroscopy Mission (XRISM), launched into low-Earth orbit in 2023, observes the reflection of solar flare X-rays in the Earth's atmosphere as a by-product of celestial observations. Using a $\sim$one-year data set covering from October 2023 to November 2024, we report on our first results of the measurement of the metal abundance pattern and high-resolution Fe-K spectroscopy.
%
The abundances of Mg, Si, S, Ar, Ca, and Fe measured with the CCD detector Xtend during M- and X-class flares show the inverse-first-ionization-potential (inverse-FIP) effect, which is consistent with the results of \cite{katsuda20} using the Suzaku satellite. The abundances of Si, S, and Ar are found to decrease with increasing flare magnitude, which is consistent with the theoretical model by Laming \citep{laming21}, whereas Ca exhibits an opposite trend. The large effective area and field of view of Xtend allow us to trace the evolution of the abundances in several X-class flare loops on a timescale of a few 100~s, finding an enrichment of low-FIP elements before flare peaks.
The high-resolution Fe-K spectrum obtained with the microcalorimeter Resolve successfully separates the Rayleigh- and Compton-scattered Fe \emissiontype{XXIV}/\emissiontype{XXV} lines and neutral or low-ionized Fe-K$\alpha$ lines. The neutral/low-ionized Fe-K$\alpha$ equivalent width shows an anti-correlation with hard X-ray flux with the best-fit power-law slope of $-0.14 \pm 0.09$, suggesting that hard X-rays from flare loops are stimulating the Fe K$\alpha$ fluorescence.
This work demonstrates that XRISM can be a powerful tool in the field of solar physics, offering valuable high-statistic CCD data and high-resolution microcalorimeter spectra in the energy range extending to the Fe-K band.

\end{abstract}

\section{Introduction}
Despite extensive studies of solar flares over half a century, key questions remain unresolved, including the physical mechanism governing the behavior of metal abundances and emission mechanism of neutral and low-ionized Fe K$\alpha$ lines.
Solar X-ray spectroscopy has been conducted with many instruments, including satellites in geostationary and low-Earth orbits and moon landing missions \citep{sylwester84, sylwester22, dennis15, vadawale21, katsuda20, mondal21, warren14, woods17, mithun22, nama23}. Metal abundances in the solar corona are known to show enhanced values of low first-ionization-potential (FIP) elements (e.g., Mg, Si, and Fe) compared with the photospheric values \citep{feldman92, dennis15}.
In active flaring regions, however, low-FIP elements are depleted rather than enhanced \citep{doschek19, brooks15, katsuda20}.
By tracing the time evolution of abundance patterns, several works have confirmed that pre-flare coronal (low-FIP enhanced) abundances get close to the photospheric (low-FIP depleted) abundance pattern during a flare, and then slowly recover the coronal values \citep{warren14, katsuda20, mondal21, nama23}.
The most plausible theoretical model is believed to be the one of Laming \citep{laming04, laming12, laming15, laming21}, where a ponderomotive force associated with Alfv\'en waves or fast-mode waves give an enhancement or depletion of low-FIP elements. This model successfully explains the abundance patterns including low-FIP elements such as Ca, Si, and S, and high-FIP elements such as Ar measured by \cite{dennis15} and \cite{katsuda20}.
X-ray observations also suggest that there is a significant flare-to-flare variation in the abundance patterns (e.g., \citealt{katsuda20, sylwester22}), which may depend on flare magnitude but observational evidence is yet to be sufficient.

Another controversial aspect of solar flares is neutral and low-ionized Fe K$\alpha$ lines \citep{neupert67, neupert71, doschek71, feldman80, tanaka84, tanaka85, zarro92, feldman96, phillips04}. The spectral shape and centroid of the Fe K$\alpha$ seem to vary depending on flares. The mechanism to stimulate these fluorescence lines is still under debate, but two candidate scenarios have been proposed: irradiation of hard X-ray photons from flare loops \citep[e.g.,][]{tanaka84} and collisions of non-thermal electrons accelerated in flare loops \citep[e.g.,][]{zarro92}. A recent X-ray study by \cite{inoue25} suggested that the former scenario is promising to explain similar neutral or low-ionized Fe K$\alpha$ lines seen in RS Canum Venaticorum type stars based on their detection of an anti-correlation between the Fe K$\alpha$ equivalent width and hard X-ray flux, which is expected only if photons are stimulating the Fe K$\alpha$ emission \citep{bai79}. In either scenario, the behavior of the Fe K$\alpha$ lines, likely depending on flares, will provide unique information of the particle-acceleration physics in flare loops.

The X-ray Imaging and Spectroscopy Mission (XRISM) is an X-ray astronomical satellite launched in September 2023 \citep{tashiro18, tashiro20, tashiro24}. XRISM is a satellite in low-Earth orbit equipped with the CCD detector Xtend (\citealt{mori23, mori24, noda25pasj, uchida25pasj}), which has a large ``grasp'' (field of view times effective area), and high-resolution spectrometer Resolve \citep{ishisaki18}.
XRISM observes the reflection of solar flare X-rays in the Earth's atmosphere during day-Earth occultation periods as a by-product of celestial observations.
Therefore, XRISM offers good capability of the spatially unresolved spectroscopic study of reflected solar flare X-rays, in a similar manner to the previous study using the Suzaku X-ray satellite \citep{katsuda20}, which was similarly in low-Earth orbit.

This work aims to present our initial results on solar flares using a one-year data set collected around the solar maximum. Section~\ref{sec-obs} describes the data set and data reduction. Section~\ref{sec-analysis} summarizes our analysis and results on Fe-K$\alpha$ emission and abundance measurement. We discuss the origin of the Fe-K$\alpha$ emission and abundance pattern and its dependence on flare magnitude in Section~\ref{sec-discussion}. Section~\ref{sec-conclusion} summarizes our findings and interpretation.

\begin{figure*}[htb!]
    \centering
    \includegraphics[width=15cm]{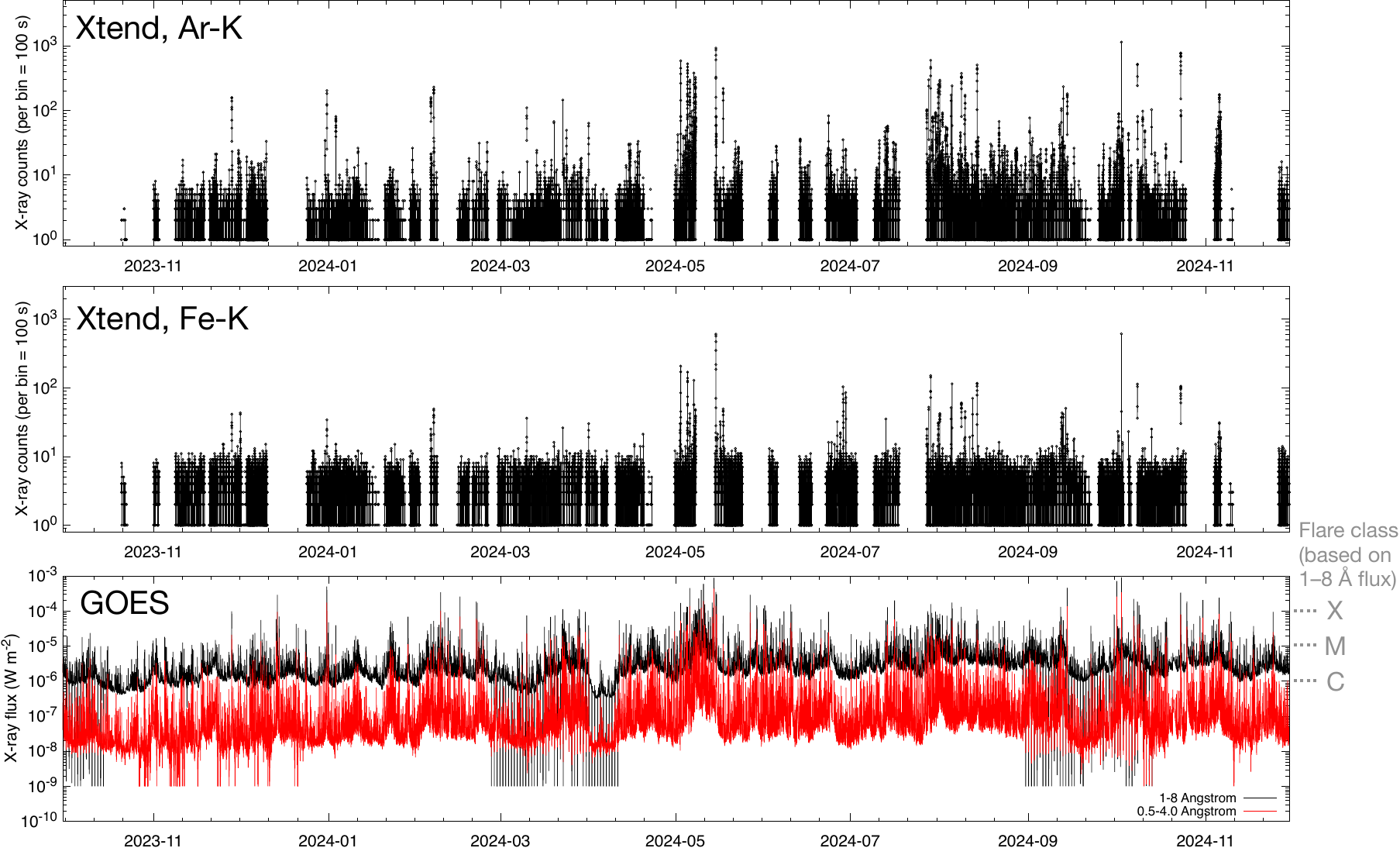}
    \flushleft
    \caption{XRISM Xtend light curves of Ar-K (2.85--3.15~keV) and Fe-K (6--7~keV) energy bands extracted from day-Earth {occultation} periods and GOES light curves. The data of the CCD3 and CCD4 are used for the Xtend light curves.
    Alt text: {Light curves of Argon K and Iron K count rates measured with Xtend and fluxes by GOES from October 2023 to December 2024.}}
    \label{fig-lc}
\end{figure*}

\begin{figure*}[htb!]
    \centering
    \includegraphics[width=15cm]{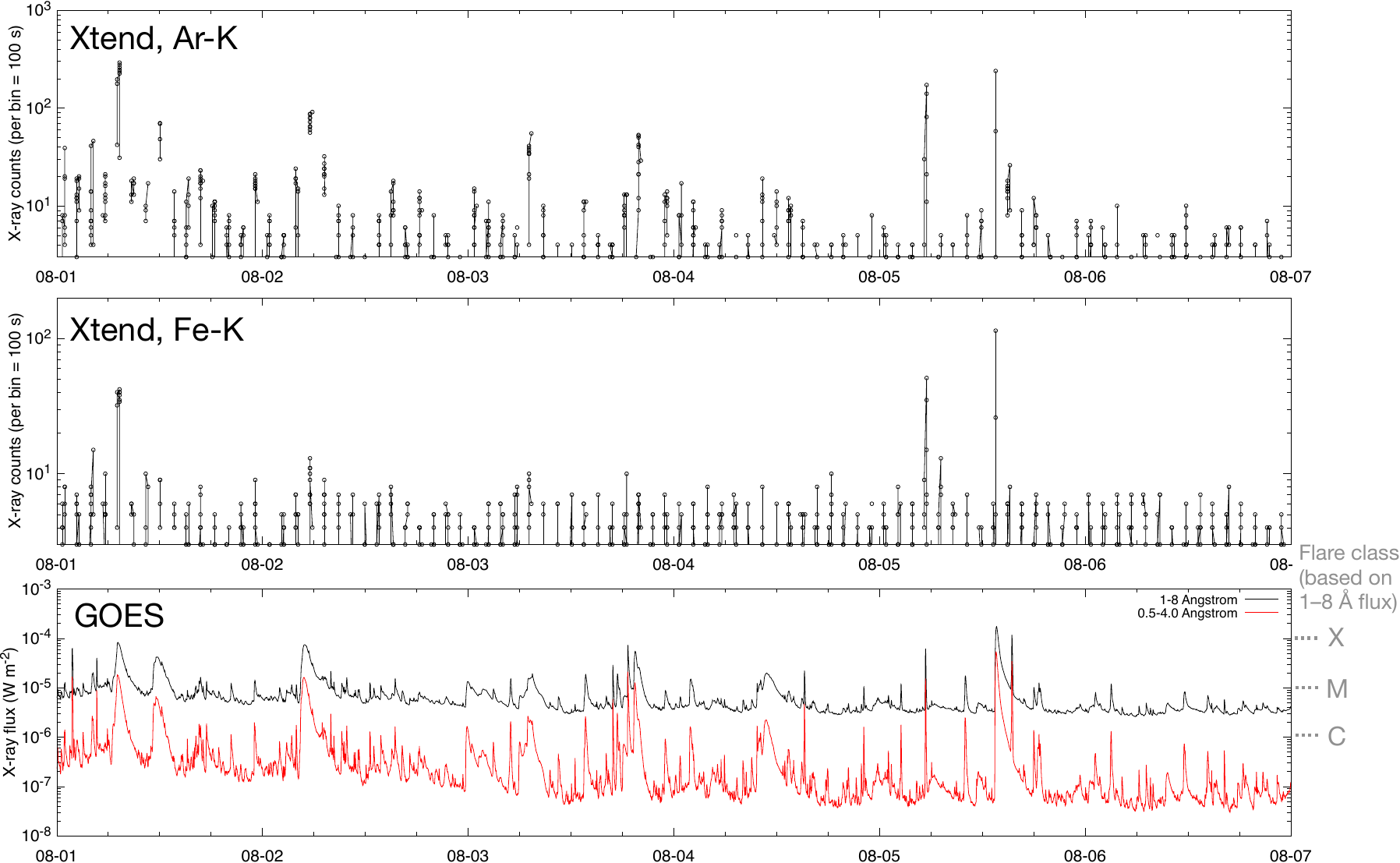}
    \flushleft
    \caption{{Same as Figure~\ref{fig-lc} but for a short period in early August 2024.
    Alt text: Same light curves as those in Figure~\ref{fig-lc} but from August 1st to 7th.}}
    \label{fig-lc2}
\end{figure*}

\section{Observations and data reduction}\label{sec-obs}
We use {the day-Earth occultation data} both with Xtend and Resolve onboard XRISM.
We use the reprocessed data stored in the XRISM ``Trend archive'' (Rev.~3)\footnote{\url{https://data.darts.isas.jaxa.jp/pub/xrism/data/trend/rev3/}}, which have been made with a screening criteria on the elevation angle, ``ELV<0 \&\& NTE\_ELV>0''\footnote{{The ELV and NTE\_ELV represent elevation angles from the Earth and night Earth limbs, respectively.}}.
For Xtend, we use observations starting from the activation of Xtend, October 21, 2023 \citep{suzuki25jatis}, to November 31, 2024. {We exclude observations during November 11--23, 2024, in which data from CCD3 and 4 were temporarily unavailable and the performance of CCD1 and 2 were degraded.\footnote{\url{https://xrism.isas.jaxa.jp/research/observers/operation_log/Xtend/index.html}}} 
Xtend has four observation modes, full-window, {full-window}+0.1-s burst, 1/8-window, and 1/8-window+{0.1-s} burst modes. The full-window mode applies the same CCD clocking for all four CCD chips with a frame exposure of 3.96~s, which is the same for CCD3 and 4 in the other observation modes, whereas CCD1 and 2 are operated with different clocking modes.
Since the observation mode for each celestial observation is determined based on the associated scientific requirement, the day-Earth {occultation data} are a Chimera of multiple observation modes.
Considering this situation, we make three data sets from the day-Earth {occultation data} for Xtend data analysis: (1) all four CCDs in the full-window observations (for images), (2) CCD3 and 4 in all the observation modes (for light curves), and (3) a combination of the former two data sets (for spectra).
For Resolve, we use observations where the nominal operation was conducted, December 25, 2023 to November 31, 2024. We use the entire array excluding the calibration pixel (pixel No. 12) and {a pixel with unusual gain variation characteristics (pixel No. 27)}. We exclude some observations with anomalies or in which special calibration activities were operated: ObsIDs 300036010, 100011010, 100011011, 100011012, 201049010, 201103010, and 201131010.\footnote{\url{https://xrism.isas.jaxa.jp/research/observers/operation_log//index.html}}

In addition to the standard screening done for the trend archive data, we apply a screening to reduce the particle-induced background and only use the {high-resolution primary (Grade 0)} events for Resolve
and 
to remove the two leading and trailing rows of the charge-injection rows and regions affected by the calibration sources (DETY>360 \&\& DETY<1440) for Xtend\footnote{Following XRISM Quick Start guide v2.3: \url{https://heasarc.gsfc.nasa.gov/docs/xrism/analysis/quickstart/index.html} and \cite{nakajima18}.}. {The Grade 0 events of Resolve are those with the best energy resolution.}
The above ``DETY'' screening on Xtend is applied to light curves and spectra but not to images.

We use HEASoft v6.34 \citep{heasarc14} for the data processing. In the spectral analysis, we use XSPEC v12.14.1 \citep{arnaud96} with the $C$-statistic \citep{cash79}, and AtomDB v3.0.9 \citep{foster12}.
Redistribution matrix files (RMFs) are generated with the tasks {\tt xtdrmf} and {\tt rslmkrmf} for Xtend and Resolve, respectively, and XRISM CALDB v20241115.
For Resolve, we use the RMF type ``XL'', where the line-spread function components include the Gaussian core, exponential tail to low energy, escape peaks, and Si fluorescence.
Auxiliary response files (ARFs) are generated with the {\tt xaarfgen} task assuming uniform emission on the plane of the sky.
Errors quoted in the text, figures, and tables indicate $1\sigma$ confidence intervals.

\begin{figure*}[htb!]
    \centering
    \includegraphics[width=16cm]{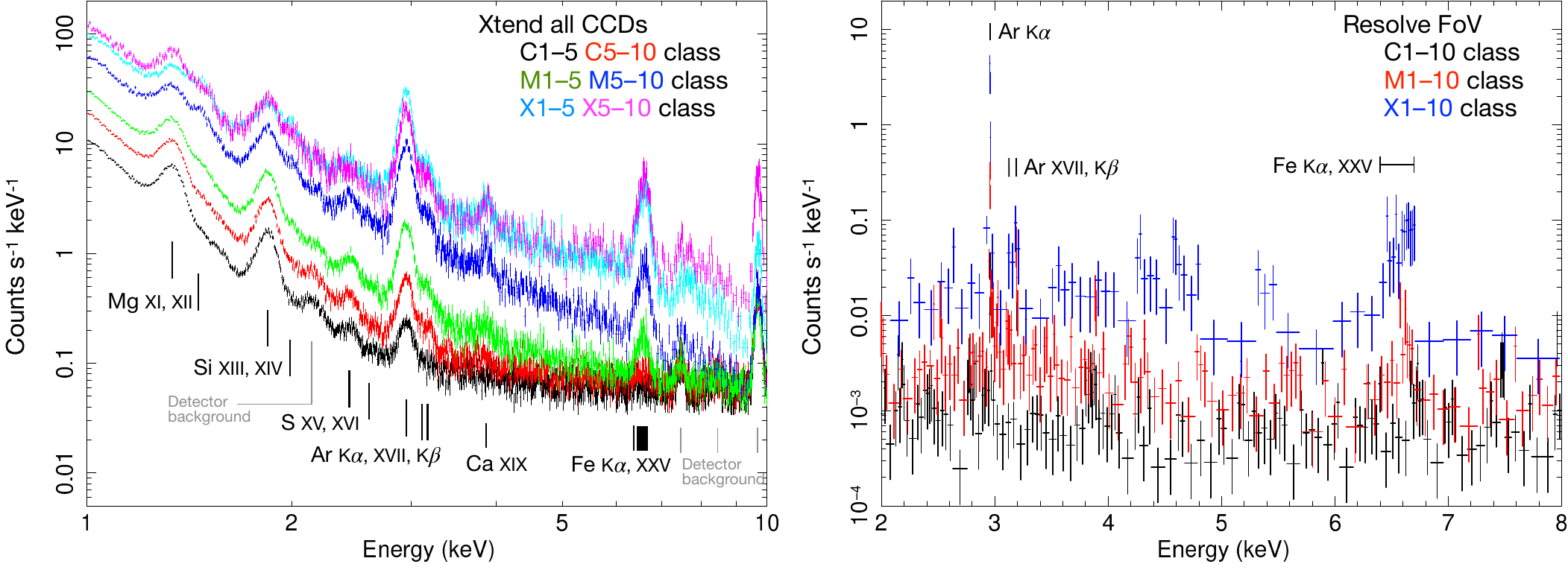}
    \flushleft
    \caption{XRISM Xtend and Resolve spectra extracted from day-Earth {occultation} periods with different flare classes.
    Alt text: {Xtend spectra (1 to 10 kilo electronvolt) for six different flare classes (left) and Resolve spectra (2 to 8~kilo electronvolt) for three different flare classes (right).}}
    \label{fig-spec}
\end{figure*}

\section{Analysis and results}\label{sec-analysis}
\subsection{Light curves}
Figure~\ref{fig-lc} shows light curves obtained with Xtend for the energy bands 2.85--3.15~keV (Ar-K) and 6.0--7.0~keV (Fe-K), and those with Geostationary Operational Environmental Satellites (GOES)\footnote{{\url{https://lasp.colorado.edu/space-weather-portal/about/latis#datasets}}}.
{Figure~\ref{fig-lc2} shows the same light curves for a short period in August 2024.}
The sharp peaks in the Xtend data mostly correspond to the flares captured by GOES.
Some minor peaks seen in the Xtend Fe-K light curve without corresponding peaks in the GOES data {(e.g., those in the end of June 2024 seen in Figure~\ref{fig-lc})} are likely due to coronal mass ejections, in which the Xtend spectra show enhanced hard X-ray continua (Kobayashi et al., PASJ, in prep.; Ishi et al., PASJ, in prep.).
The intervals without Xtend photon counts are the periods when no day-Earth {occultation data} are available. This depends on the directions toward the celestial source and the Sun.

\subsection{Spectra}\label{sec-spec}
Figure~\ref{fig-spec} shows Xtend and Resolve spectra extracted from data stacked for different flare magnitudes. 
Here we use the flare periods reported by GOES\footnote{\url{https://data.nas.nasa.gov/helio/portals/solarflares/datasources.html}} for spectral extraction.
The total numbers and exposures of flares in the Xtend spectra for different flare classes are summarized in Table~\ref{tab-obs}.
The Xtend spectra are extracted from all four CCDs in the full-window observations. We use energies above 1~keV to avoid possible contamination of the ``crosstalk'' (pseudo) events \citep{nakajima18}, which is a treatment similar to preceding work using Xtend \citep{suzuki25v4641}.
One can see clear emission lines of Mg \emissiontype{XI}, Mg \emissiontype{XII}, Si \emissiontype{XIII}, Si \emissiontype{XIV}, S \emissiontype{XV}, S \emissiontype{XVI}, Ar {K$\alpha$}, Ar {K$\beta$}, Ca \emissiontype{XIX}, and Fe {K$\alpha$} in the Xtend spectra. The Resolve spectra show clear Ar {K$\alpha$} and Ar {K$\beta$} lines in all the C, M, and X-class flares and Fe \emissiontype{XXV} lines in the M- and X-class flares, while lines from the other elements are not very clear due to much less statistics compared with Xtend.


\begin{table}[htb!]
    \caption{{Basic information of flares used in the spectral analysis of this work}}
    \centering
    \begin{tabular}{l r r}
    \hline\hline
    Flare class & Number of flares & Total exposure (ks) \\
    \hline
    C1--5 & 1944 & 98 \\
    C5--10 & 1033 & 50 \\
    M1--5 & 594 & 31 \\
    M5--10 & 68 & 5.6 \\
    X1--5 & 34 & 2.6 \\
    X5--10 & 5 & 0.9 \\
    \hline
    \end{tabular}
    \label{tab-obs}
\end{table}

\begin{figure*}[htb!]
    \centering
    \includegraphics[width=16cm]{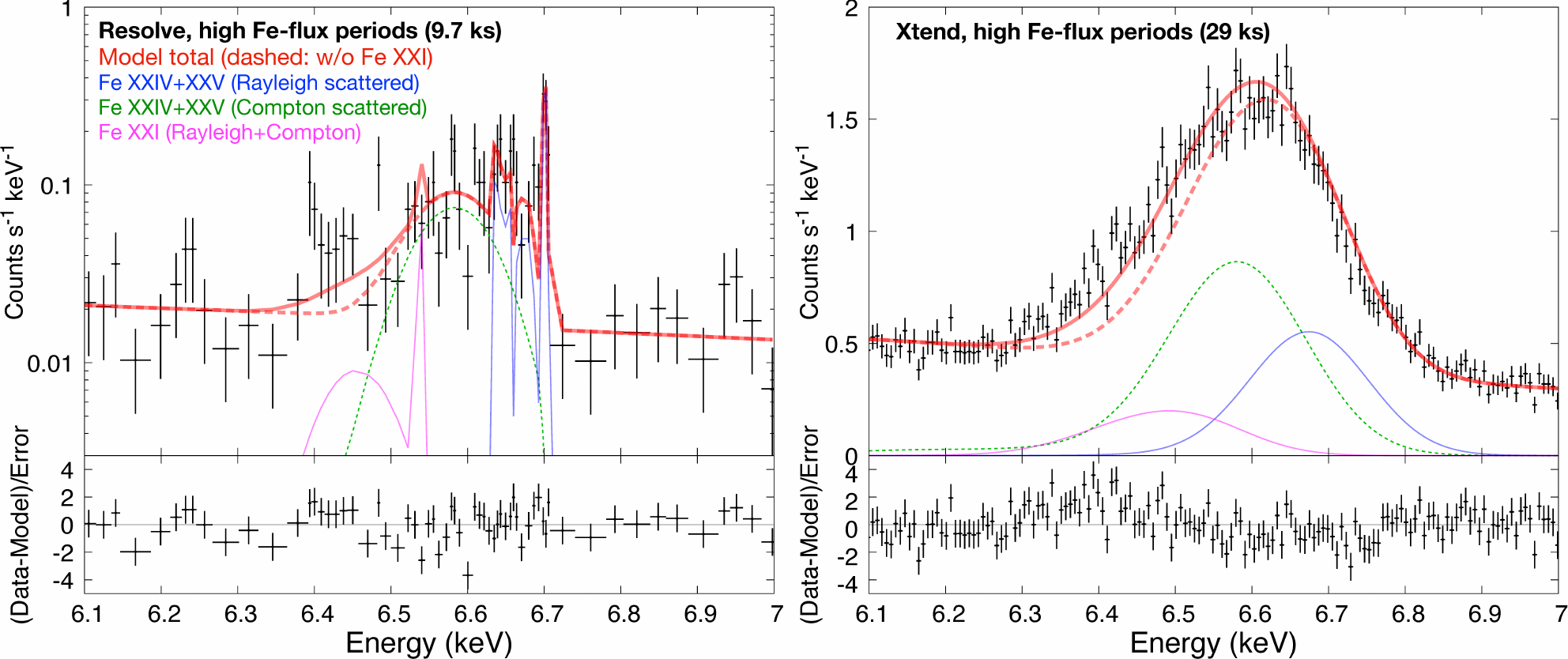}
    \flushleft
    \caption{Resolve and Xtend spectra extracted from day-Earth {occultation} periods with bright Fe-K fluxes and best-fit physical spectral models. The spectral models include Rayleigh- and Compton-scattered Fe \emissiontype{XXIV}, \emissiontype{XXV}, and \emissiontype{XXI} lines. {The lower panels show the residuals between the data and the model total.}
    Alt text: {Resolve (left) and Xtend (right) spectra of the Iron-K energy band with the three model components, their total, and the total without Iron XXI component. Both are shown for 6.1 to 7.0 kilo electronvolt.}}
    \label{fig-spec2}
\end{figure*}

\begin{figure}[htb!]
    \centering
    \includegraphics[width=8cm]{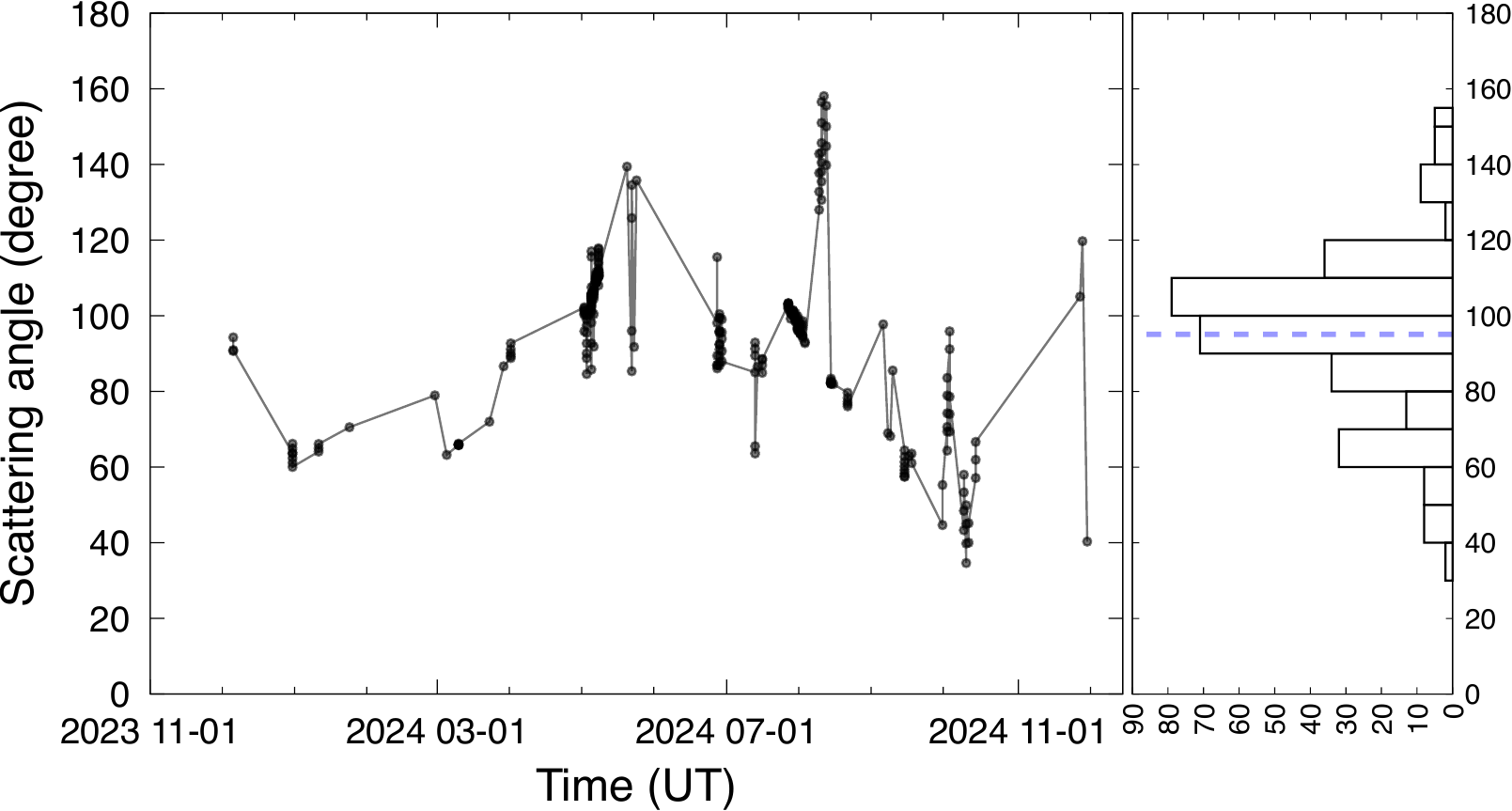}
    \flushleft
    \caption{Scattering angle of solar X-rays observed with XRISM as a function of time (left) and a histogram of the scattering angle (right). The angles are calculated for periods in which bright Fe-K emission is observed with Xtend. The blue dashed line in the right panel indicates the mean scattering angle.
    Alt text: Scattering angles shown from November 2023 to December 2024 (left) and their histogram (right).}
    \label{fig-angle}
\end{figure}

\begin{figure*}[htb!]
    \centering
    \includegraphics[width=16cm]{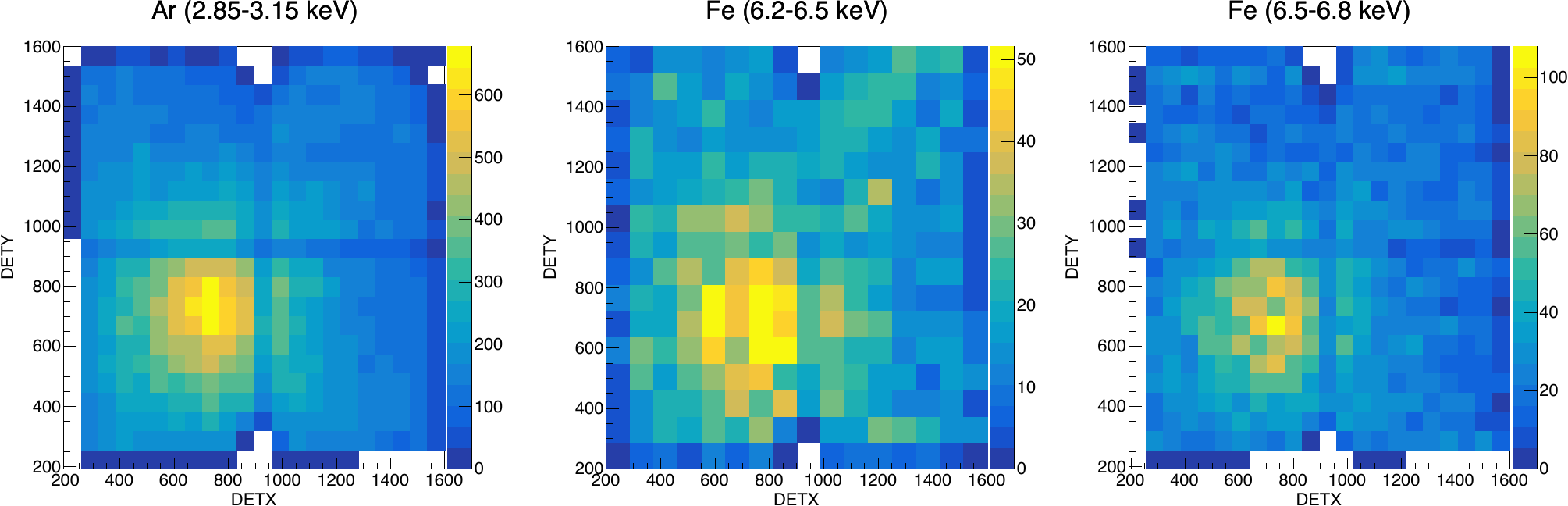}
    \flushleft
    \caption{Xtend images for the three energy bands extracted from day-Earth {occultation} periods with bright Fe-K fluxes. The aim point is located at (DETX, DETY) = (734.12, 730.85).
    Alt text: {Three images from 2.85 to 3.15~kilo electronvolt, 6.2 to 6.5~kilo electronvolt, and 6.5 to 6.8~kilo electronvolt, all of which similarly show peaks at around the aim point.}}
    \label{fig-image}
\end{figure*}

\begin{figure}[htb!]
    \centering
    \includegraphics[width=7cm]{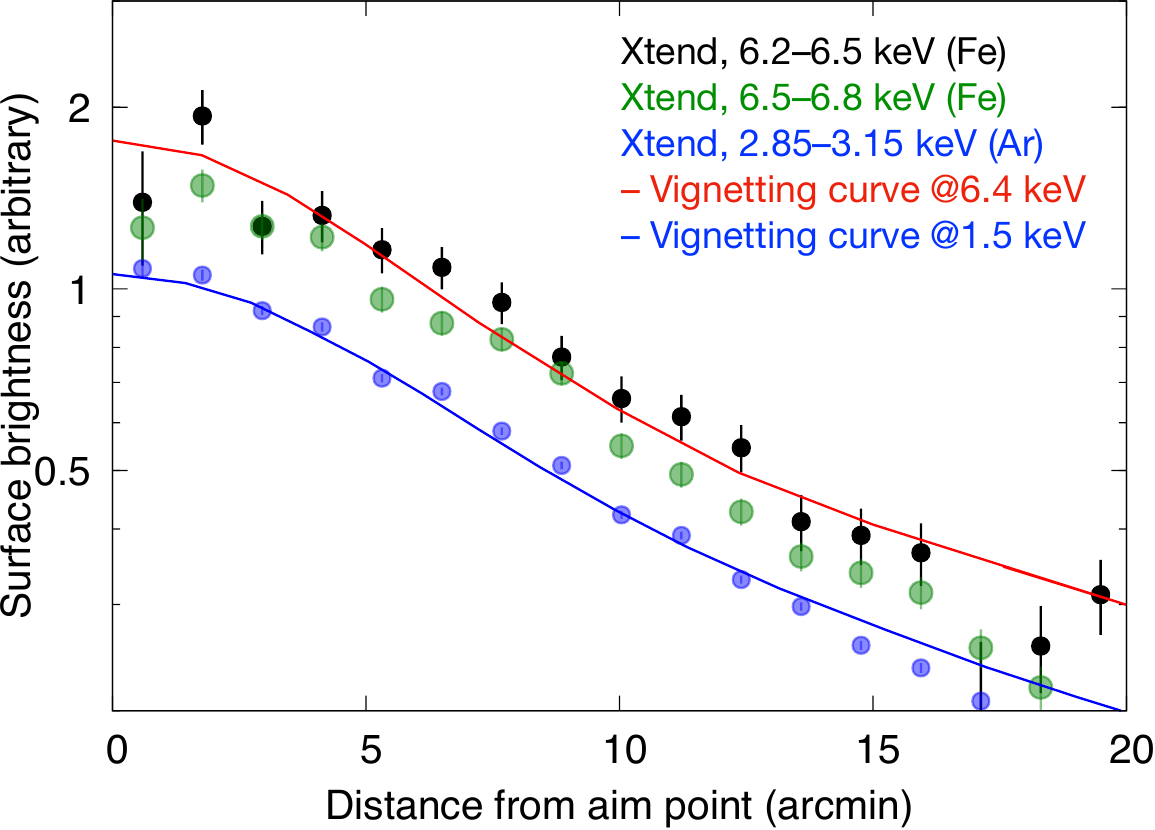}
    \flushleft
    \caption{Radial profiles of X-ray fluxes extracted from three energy ranges and the vignetting curves obtained in on-ground experiments.
    Alt text: {Radial profiles from three energy ranges, 2.85 to 3.15~kilo electronvolt, 6.2 to 6.5~kilo electronvolt, and 6.5 to 6.8~kilo electronvolt, and two reference vignetting curves at 1.5~kilo electronvolt and 6.4~kilo electronvolt, shown for 0 to 20 arcminute.}}
    \label{fig-radial}
\end{figure}

\subsubsection{Fe-K energy band}\label{sec-fe}

\begin{figure}[htb!]
    \centering
    \includegraphics[width=8cm]{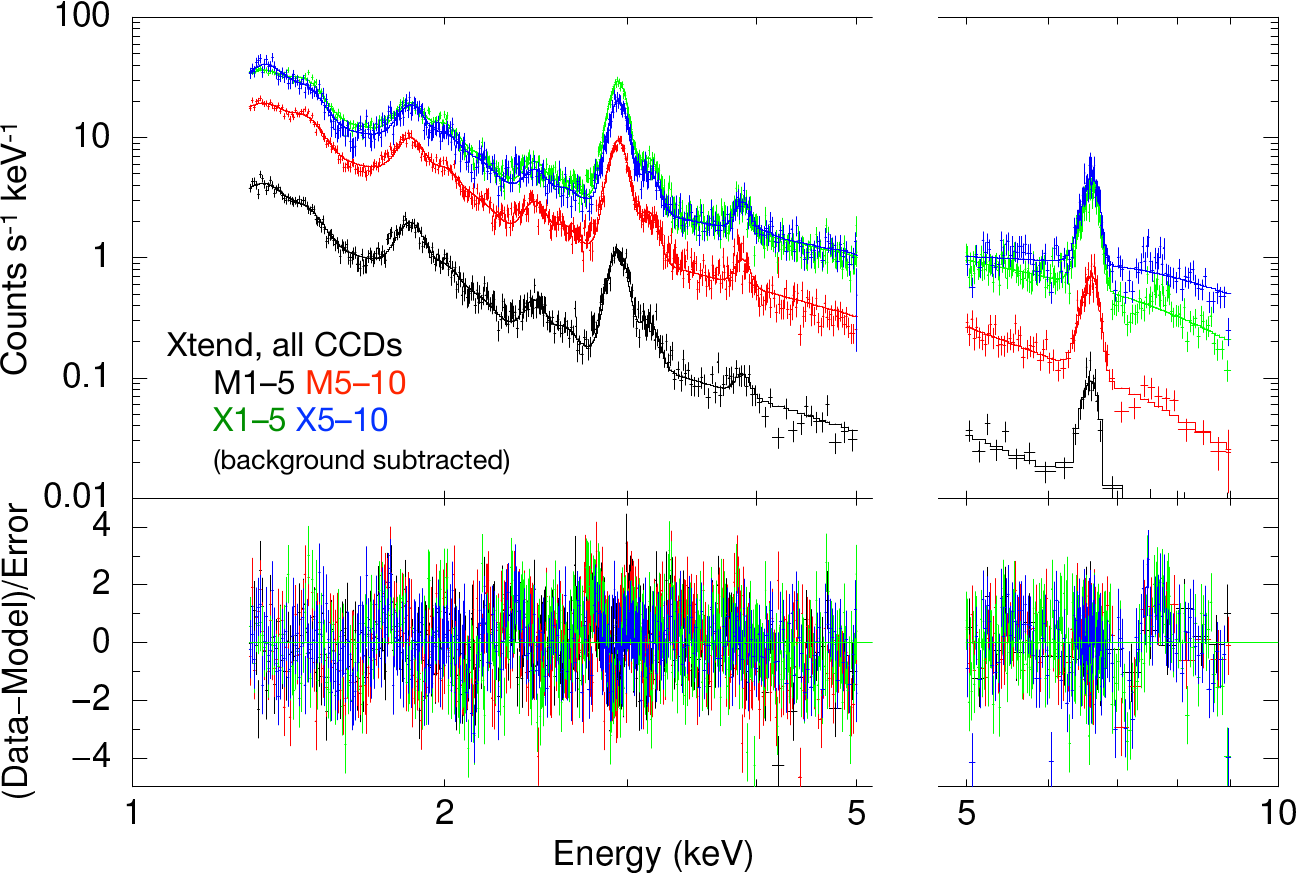}
    \flushleft
    \caption{Xtend spectra extracted from day-Earth periods with different flare magnitudes and best-fit phenomenological spectral models. The pre-flare background spectra are subtracted from the Xtend spectra. {The gap at 5~keV is intentional, indicating that we separately model the 1.3--5.0~keV and 5.0--9.0~keV energy ranges.}
    Alt text: Four Xtend spectra extracted from four different flare classes and models.}
    \label{fig-spec_abund}
\end{figure}

\begin{figure}[htb!]
    \centering
    \includegraphics[width=8cm]{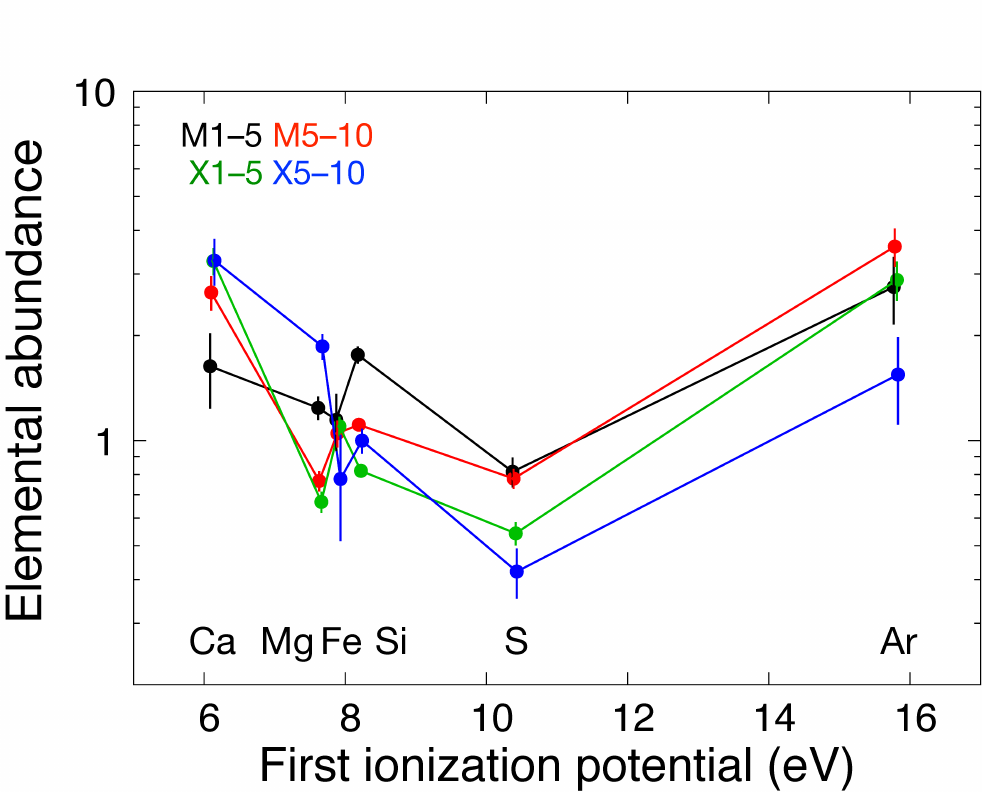}
    \flushleft
    \caption{Metal abundances relative to the photospheric values derived based on the Xtend spectra, shown against the first ionization potential.
    Alt text: {Metal abundances of Calcium, Magnesium, Iron, Silicon, Sulfur, and Argon over first ionization potential in 5 to 17~electronvolt, shown for four different flare classes.}}
    \label{fig-abund}
\end{figure}

\begin{figure*}[htb!]
    \centering
    \includegraphics[width=12cm]{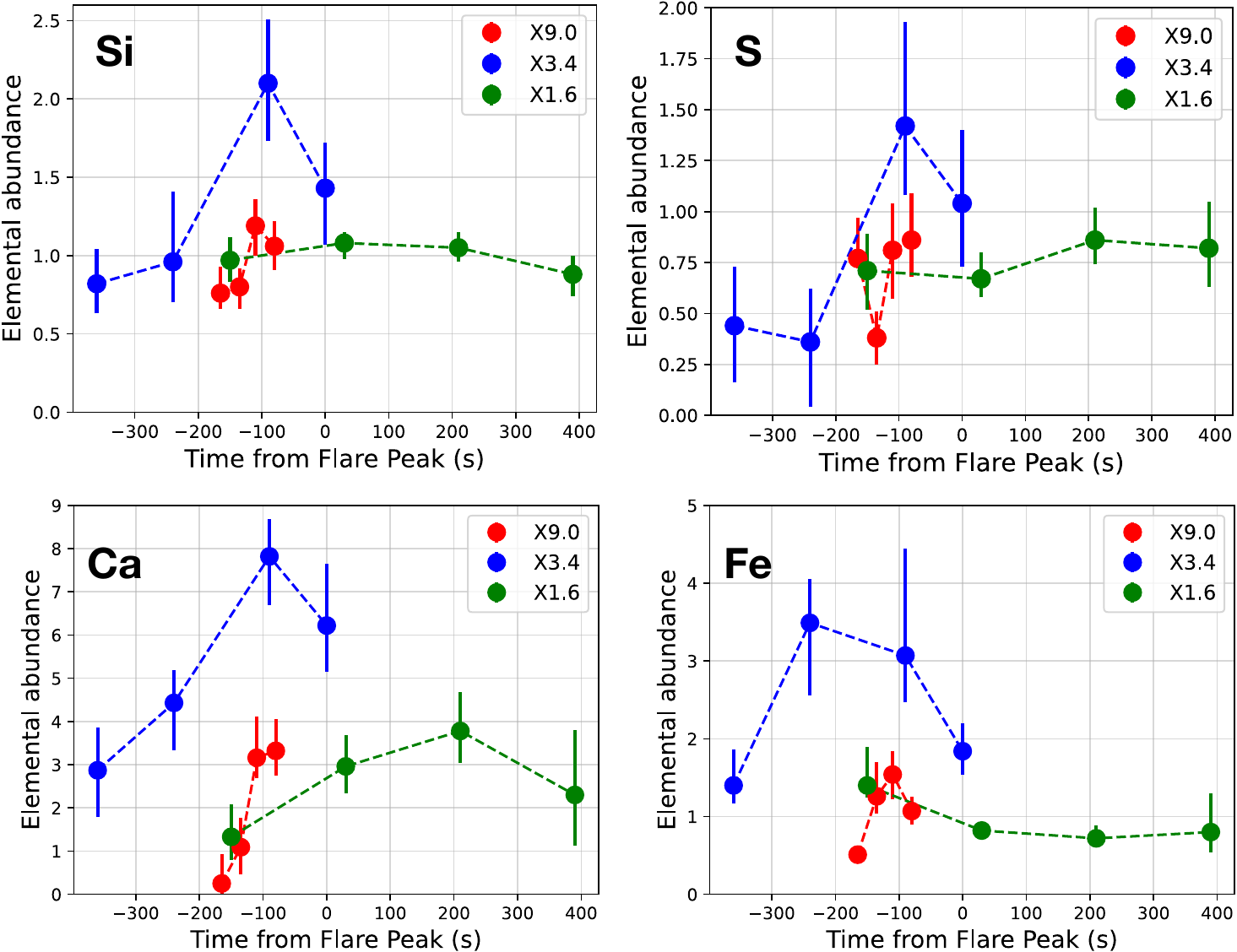}
    \flushleft
    \caption{Metal abundances with respect to the photospheric values as a function of time around the peaks of three X-class flares.
    Alt text: {Silicon, Sulfur, Calcium, and Iron abundances as a function of time, shown for three X-class flares in minus 400 to plus 400 seconds with respect to the flare peaks.}}
    \label{fig-abund2}
\end{figure*}

\begin{figure*}[htb!]
    \centering
    \includegraphics[width=16cm]{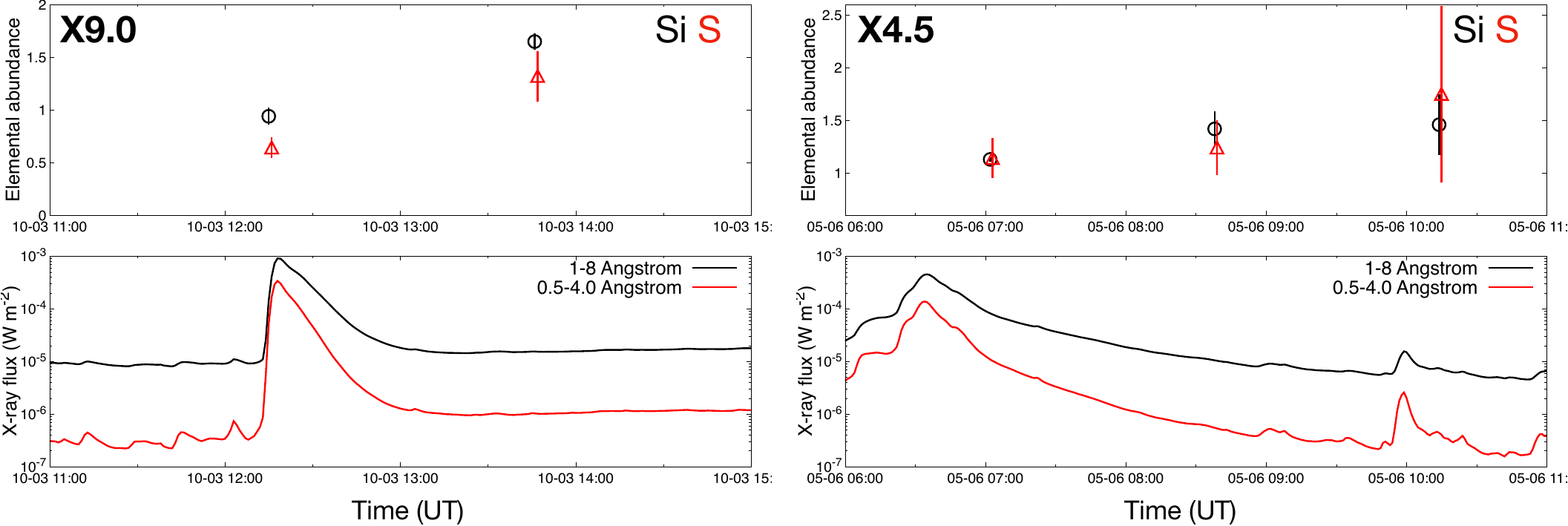}
    \flushleft
    \caption{Metal abundances with respect to the photospheric values as a function of time (upper panels) and corresponding GOES light curves (lower panels) during and after two X-class flares.
    Alt text: {Silicon and Sulfur abundances and GOES fluxes as a function of time for two X-class flares in around minus 1 to plus 3 hours with respect to the flare peaks.}}
    \label{fig-abund3}
\end{figure*}

In order to examine the Fe-K spectral shape in detail, we extract spectra from the periods when the Fe-K flux is especially high in the Xtend light curve. Figure~\ref{fig-spec2} shows the extracted spectra.
To give a physical spectral model, we consider the Fe \emissiontype{XXV}-{\it w, x, y}, and {\it z}, and Fe \emissiontype{XXIV}-{\it j} and {\it k} lines \citep{neupert67, feldman80, tanaka84} and their reflection in the Earth's atmosphere.
{The reflection in the Earth's atmosphere is calculated via the Monte Carlo method (see, e.g., \citealt{churazov08}). There, the input solar X-ray emission is assumed to be a collisional ionization equilibrium plasma with a temperature of 1.5~keV (i.e., the {\tt apec} model in XSPEC), and the atmospheric composition is assumed to be N, O, and Ar with a number fraction of 0.781, 0.209, and 0.0093, respectively. {The temperature of 1.5~keV is consistent with those of the X-class flares included in our study determined based on the GOES fluxes ($1.4\pm0.2$~keV; see Appendix~\ref{sec-alpha}).}
The output albedo emission is then estimated by collecting reflected photons at a scattering angle of around 90$^\circ$, which is roughly consistent with the averaged reflection angles in the day-Earth data used in this work ($\approx 95^\circ$; Figure~\ref{fig-angle}).}
{We note that the temperature and scattering angle can be better constrained by the observed spectra in principle. Given that our aim is not to determine these parameters, which also requires non-negligible computational costs, we adopt the temperature of 1.5~keV and scattering angle of 90$^\circ$ as a representative parameter set to explain observations. The systematics due to the uncertainties of these parameters are discussed later in this section and in Section~\ref{sec-ana-abund}.
}

The resultant prediction of the Rayleigh- and Compton-scattered lines are shown in Figure~\ref{fig-spec2}.
{The Rayleigh-to-Compton flux ratio obtained from our calculation is $\approx 0.543$.}
One can see that both spectra in 6.5-6.8~keV are explained well.
However, there remains some excess in 6.4--6.5~keV (red dashed lines in Figure~\ref{fig-spec2}). This should indicate the presence of neutral or low-ionized Fe K$\alpha$. The neutral Fe K$\alpha$ at $\approx 6.40$~keV would be observed as a combination of narrow $\approx 6.40$~keV (Rayleigh) and broad $\approx 6.32$~keV (Compton) peaks, and thus would not be able to explain the observed excess. Thus, we instead add a low-ionized Fe \emissiontype{XXI} K$\alpha$ line at $\approx 6.537$~keV, which is a relatively prominent line below the Fe \emissiontype{XXV} energy band (e.g., \citealt{feldman80, tanaka85}). This additional model is shown in Figure~\ref{fig-spec2}. This in particular improves the model for the Xtend spectrum. With more statistics especially with Resolve, we will be able to distinguish more confidently which charge states of Fe contribute to the emission in $\approx 6.4$--6.5~keV.
{Also, the left wing of the Compton-scattered Fe \emissiontype{XXIV} and \emissiontype{XXV} lines should have complex structures including additional peaks produced via Compton ionization of N and O atoms, which may be resolved with better statistics.}

{While a variation of the flare temperature within $1.4\pm0.2$~keV would not affect the Fe-K spectral modeling, the uncertainties in the scattering angle ($\sim 95\pm20^\circ$ from Figure~\ref{fig-angle}) would alter the flux ratios of the Rayleigh- and Compton-scattered photons by $\lesssim 30\%$, and the peak energy of the Compton-scattered photons by $\lesssim 30$~eV. These would not be significant for our main conclusion here, which is the fact that the XRISM Fe-K spectrum can be reasonably understood with a combination of highly-ionized and neutral/low-ionized Fe emission.}

{In order to confirm that our XRISM data predominantly consist of reflected solar X-rays, we here examine images observed with Xtend.}
If photons are focused (come through mirrors), observed images would show a peak at the aim point and flux decrease toward off-axis angle, which is described with the vignetting curve. On the other hand, we expect images with a constant count rate over the entire detector area if photons are {not coming through mirrors}.
Figure~\ref{fig-image} shows Xtend images for the three energy bands, neutral Ar-K$\alpha$ and Fe-K$\alpha$, and Fe \emissiontype{XXV} lines ($\approx$6.6~keV) extracted from the periods with bright Fe-K emission. All three images show peaks coincident with the aim point and a gradual decrease with increasing off-axis angle. This is consistent with our expectation that these emission lines mostly originate from the Earth's atmosphere and we observe flat-field emission over the field of view.
In order to be more quantitative, we compare the observations with the vignetting curves (effective areas as a function of off-axis angle) by creating corresponding radial profiles shown in Figure~\ref{fig-radial}. Here we show the vignetting curve measured in on-ground experiments and the observed radial profiles after subtracting the detector background levels. One can see that all three energy bands can be well explained with the vignetting curves. This result confirms that these emission lines can be described with a flat-field emission almost without contamination from other origins, e.g., enhanced detector background observed several times with Xtend (Kobayashi et al., PASJ, in prep.; Ishi et al., PASJ, in prep.).

\subsection{Abundance measurement of flare loops}\label{sec-ana-abund}

\begin{table*}[htb!]
    \caption{Best-fit parameters to explain the X-ray spectra in four different flare classes}
    \centering
    \begin{tabular}{l r r r r}
        \hline\hline 
        Parameter & M1--5 & M5--10 & X1--5 & X5--10 \\
        \hline
Mg \emissiontype{XI} (eV)\footnotemark[${*}$] & $81 \pm 6$ & $54 \pm 4$ & $48 \pm 3$ & $105 \pm 8$  \\
Si \emissiontype{XIII} (eV)\footnotemark[${*}$]  & $206 \pm 11$ & $151 \pm 6$ & $119 \pm 5$ & $139 \pm 11$  \\
S \emissiontype{XV} (eV)\footnotemark[${*}$]  & $118 \pm 11$ & $114 \pm 7$ & $84 \pm 6$ & $68 \pm 11$  \\
Ar \emissiontype{XVII} (eV)\footnotemark[${*}$]  & $117 \pm 26$ & $136 \pm 17$ & $120 \pm 16$ & $86 \pm 24$  \\
Ca \emissiontype{XIX} (eV)\footnotemark[${*}$]  & $80 \pm 19$ & $104 \pm 12$ & $115 \pm 10$ & $115 \pm 17$  \\
Fe (6.675~keV) (eV)\footnotemark[${*}$]  & $1344 \pm 153$ & $1253 \pm 80$ & $1300 \pm 48$ & $977 \pm 57$  \\
Fe (6.537~keV) (eV)\footnotemark[${*}$]  & $658 \pm 608$ & $406 \pm 160$ & $389 \pm 82$ & $248 \pm 112$  \\
Si \emissiontype{XIV}--\emissiontype{XIII} flux ratio  & $0.202 \pm 0.015$ & $0.308 \pm 0.007$ & $0.448 \pm 0.006$ & $0.286 \pm 0.028$   \\
{$\alpha$ of DEM model} & $-0.85 \pm 0.06$ & $-0.398 \pm 0.009$ & $0.0352 \pm 0.0005$ & $-0.48 \pm 0.05$   \\
Power-law norm. (0.5--5.0~keV)\footnotemark[${\dag}$] & $5.77 \pm 0.21$ & $27.01 \pm 0.63$ & $47.43 \pm 1.00$ & $38.28 \pm 1.52$  \\
Power-law  index (0.5--5.0~keV) & $3.01 \pm 0.04$ & $2.61 \pm 0.03$ & $2.25 \pm 0.02$ & $2.09 \pm 0.04$  \\
Power-law norm. (5.0--8.0~keV)\footnotemark[${\dag}$] & $11.3 \pm 10.0$ & $56.4 \pm 19.7$ & $21.2 \pm 3.68$ & $2.31 \pm 0.53$   \\
Power-law index (5.0--8.0~keV) & $3.47 \pm 0.51$ & $3.18 \pm 0.20$ & $1.78 \pm 0.09$ & $0.36 \pm 0.12$   \\
\hline 
    \end{tabular}
    \label{tab-fit}
    \begin{tabnote}
    \flushleft
    \footnotemark[${*}$] Equivalent width in units of eV.\\
    \footnotemark[${\dag}$] Normalization in units of photons~keV$^{-1}$~cm$^{-2}$~s$^{-1}$ at 1~keV.
    \end{tabnote}
\end{table*}

\begin{table}[htb!]
    \caption{Metal abundances relative to the photospheric values \citep{lodders03} measured for four different flare classes}
    \centering
    \begin{tabular}{l r r r r}
        \hline\hline 
        Element & M1--5 & M5--10 & X1--5 & X5--10 \\
        \hline
Mg & $1.24 \pm 0.10$ & $0.77 \pm 0.05$ & $0.67 \pm 0.05$ & $1.86 \pm 0.16$ \\
Si & $1.76 \pm 0.10$ & $1.11 \pm 0.05$ & $0.82 \pm 0.04$ & $1.00 \pm 0.08$ \\
S & $0.81 \pm 0.08$ & $0.78 \pm 0.05$ & $0.54 \pm 0.04$ & $0.42 \pm 0.07$ \\
Ar & $2.75 \pm 0.60$ & $3.59 \pm 0.45$ & $2.88 \pm 0.37$ & $1.54 \pm 0.43$ \\
Ca & $1.63 \pm 0.40$ & $2.65 \pm 0.30$ & $3.27 \pm 0.29$ & $3.28 \pm 0.50$ \\
Fe & $1.15 \pm 0.21$ & $1.05 \pm 0.09$ & $1.10 \pm 0.07$ & $0.78 \pm 0.26$ \\
\hline 
    \end{tabular}
    \label{tab-abund}
\end{table}

Using the Xtend spectra for different flare magnitudes, we here measure the metal abundance pattern and its dependence on the flare magnitude. Because day-Earth data provide spatially unresolved spectra, it is hard to extract flaring components from non-flaring background emission for small flares. Thus we focus on flare magnitudes higher than M1 in our abundance measurement.
In order to measure metal abundances from the reflected solar-flare spectra, we adopt the methodology introduced by \cite{katsuda20}: converting measured equivalent widths {(line-to-continuum ratios)} to abundances assuming a multi-temperature optically-thin thermal plasma model for the solar flare emission before it is deformed by the reflection in the Earth's atmosphere.
{The continuum emission is defined as the sum of free–free and free–bound emissions. The representative electron temperature is determined by the flux ratio of the Si \emissiontype{XIV} and \emissiontype{XIII} lines. Elemental abundances are then obtained by comparing observed and theoretical line-to-continuum ratios.
}

{We adopt {a plasma model with the differential emission measure (DEM) following a power-law in temperature (hereafter, power-law DEM model)} as a general emission model to describe the flare spectra (e.g, \citealt{caspi14, kepa18}).
Note that \cite{katsuda20} compared the results using the {power-law} DEM and two-temperature models, finding similar abundance measurement results, i.e., they are consistent within $\lesssim 20$\%.}
For the {power-law} DEM model, we use the XSPEC model {\tt cevmkl} with the {minimum and} maximum electron {temperatures $kT_{\rm min} \approx 0.03$~keV and} $kT_{\rm max} = 3.5$~keV, {respectively}. We simulate mock spectra based on the {power-law} DEM model with a variety of the parameter $\alpha$, which determines how EMs are weighted by electron temperature $kT_{\rm e}$ as ${\rm EM} {\propto} ({kT_{\rm e})^\alpha}$, and metal abundance, and evaluate the equivalent widths of the lines using the same spectral model as the one used for the observations. The $\alpha$ value that best represents the observed spectrum is determined based on the observed Si \emissiontype{XIV}-to-\emissiontype{XIII} flux ratio. In this way, we can relate the observed equivalent widths to the abundances.

{In principle, it is possible to infer the actual temperature distribution from the spectral shapes obtained with Xtend. However, the spectral deformation by the reflection in the Earth's atmosphere, while unchanging the equivalent widths of line emission, makes straightforward modeling difficult and would require analytical or numerical calculations to reconstruct the original spectral shapes. As the adopted power-law DEM model explains our observations well as we see below, we here approximate the flare spectra with a single power-law DEM model.
We note that our approximation with a power-law DEM model would also be suitable considering the fact that we use the data set in which multiple flares are stacked.}

Here we evaluate the abundance patterns for different flare magnitudes by extracting data sets from the corresponding flaring periods defined by GOES.
The spectral model to evaluate the equivalent widths consists of the Mg \emissiontype{XI} and \emissiontype{XII}, Si \emissiontype{XIII} and \emissiontype{XIV}, S \emissiontype{XV} and \emissiontype{XVI}, Ar \emissiontype{XVII}, Ca \emissiontype{XIX}, and Fe lines, with one power-law continuum (see Table~\ref{tab-fit} for details). For the Fe lines, we use the physical spectral model determined in Section~\ref{sec-fe}, composed of the Rayleigh- and Compton-scattered Fe \emissiontype{XXV}, \emissiontype{XXIV}, and \emissiontype{XXI} lines. We fix the relative flux of the Rayleigh and Compton components to the value determined in Section~\ref{sec-fe} because the average reflection angle, which determines the Rayleigh-to-Compton flux ratio, does not change much for different spectra extracted for different flare magnitudes.
{The power-law component mostly represents the thermal bremsstrahlung.}
The free parameters include the fluxes of the lines and power-law flux and index.
To exclude the contribution of quiescent solar X-rays not originating from the flares, we extract a pre-flare (up to 10$^4$~s ahead of a flare) spectrum for each flare and stack all the pre-flare spectra for each of the M1--5, M5--10, X1-5, and X5--10 classes to construct background spectra (see Figure~\ref{fig-bgd} in Appendix~\ref{sec-bgd}). These background spectra are then used in the framework of the $C$-statistic in XSPEC.
Since it is generally hard to explain the entire spectral continuum in {1.3--9.0}~keV with one power-law component, we separately model the spectra in {1.3}--5.0~keV (Mg to Ca) and 5--{9}~keV (Fe).
{This likely indicates that the observed spectra can be approximated with multiple DEM components. Note that we similarly model the simulated DEM spectra in the two energy bands for consistency when converting measured equivalent widths to metal abundances.}
The best-fit models and parameters are shown in Figure~\ref{fig-spec_abund} and Table~\ref{tab-fit}, respectively. The resulting abundances are listed in Table~\ref{tab-abund} and shown as a function of FIP in Figure~\ref{fig-abund}. One can see that most parameters including the metal abundances vary with the flare class.
We give a deeper look at the variation of the flare temperature depending on the flare class, which is given in Appendix~\ref{sec-alpha}.

We note that the Ar \emissiontype{XVII} lines at $\approx 3.135$~keV overlap with the fluorescent Ar K$\alpha$ and K$\beta$ at 2.95~keV and 3.191~keV, respectively {(Figure~\ref{fig-spec})}. By examining the Resolve spectrum extracted from periods with bright Fe-K emission, we confirm that the equivalent width unambiguously determined for the Ar \emissiontype{XVII} lines, $60\pm20$~eV, is consistent with the one derived for X5--10-class flares with Xtend ({Table~\ref{tab-fit}}). This is supportive evidence for our abundance measurement of Ar.

We also investigate the time variation of the abundance pattern during the flaring episodes. In addition to the long-term variability on a timescale of a few 1000~s examined in previous studies \citep{warren14, katsuda20, mondal21, nama23}, we also examine the variability on a shorter timescale, a few 100~s (similar to a previous work, \citealt{warren14}), by utilizing the large grasp of Xtend. The spectral model and treatment of the background are the same as above, but the modeling is done separately for short periods during flaring episodes for three high-statistics X-class (X9.0, X3.4, and X1.6) flares.
{Again, we here use a power-law DEM model. According to \cite{katsuda20}, the systematic uncertainties in the resultant abundances due to the uncertain temperature distribution would be $<20\%$, which is similar to the statistical uncertainties as one can see below, and thus would not affect our conclusion.}

In Figure~\ref{fig-abund2}, we show the short-term variability of the metal abundances around the flare peaks defined based on the GOES X-ray light curves. We find that the X9.0 and X3.4 flares consistently show positive peaks in abundance values at around 100~sec ahead of the flare peaks, while the behavior may differ for the X1.6 flare. This trend is most clearly seen in Ca, which exhibits a most prominent enhancement in abundance. We note that the time variation of the Ar abundance is statistically insignificant and is omitted here.
The long-term variability is shown in Figure~\ref{fig-abund3}. Because of the limited statistics and available time windows in the post-flare periods, only the X9.0 flare in above three is found to be suitable for this study. We add another X-class flare, the X4.5 flare, which is excluded in the above short-term study because the XRISM data failed to cover the flare peak, but is suitable in terms of the long-term variability. Also, due to the limited statistics after considering the pre-flare background, we can obtain meaningful results only for Si and S. The abundances of both elements show similar increasing trends with time.


{We here examine systematic uncertainties associated with geometrical configurations. Potential geometrical effects include variations in the (1) scattering angle (Figure~\ref{fig-angle}) and (2) fluctuation of atmospheric column density. The results that are potentially affected are (a) Fe-K spectral model and (b) Si \emissiontype{XIV}/\emissiontype{XIII} flux ratio used for abundance measurements. Note that equivalent widths, which are converted to abundances, are unaffected by the spectral deformations caused by these effects.
Although the (1) uncertain scattering angle would slightly alter the (a) Fe-K modeling as discussed in Section~\ref{sec-fe}, subsequent measurements of the Fe abundance and neutral/low-ionized Fe equivalent width are not significantly affected because of the limited energy resolution of Xtend. The slightly shifted Compton peak also changes the (a) best-fit Fe-K model, but the resultant systematic effect on the equivalent width estimates is found to be insignificant as well.
The effect (1) does not affect (b) Si \emissiontype{XIV}/\emissiontype{XIII} flux ratio as it is independent of the scattering angle.
As for (2), an increased atmospheric density would increase the photon interaction probability but the scattering rate would be unchanged because the penetration depth becomes smaller assuming a uniform density. Thus, it would not affect neither (a) nor (b). In principle, a spatial fluctuation of the upper atmospheric density (e.g., \citealt{miyoshi22, shinagawa24, katsuda24}) may alter the scattered spectral shape, which is an interesting effect but beyond the scope of this paper.}

\section{Discussion}\label{sec-discussion}
\subsection{Origin of the neutral and low-ionized Fe lines}
We have found a spectral excess in the 6.4--6.5~keV {(Figure~\ref{fig-spec2})}, coincident with neutral or low-ionized Fe-K$\alpha$ lines, in the Resolve and Xtend spectra extracted from Fe-K bright periods. Since the Xtend image shows a concentration around the optical axis, the neutral/low-ionized Fe-K photons are mostly coming from the field of view of Xtend. Then the candidates for their origin are fluorescence in the Earth's atmosphere and solar flares.

The Fe K$\alpha$ fluorescence lines in the Earth's atmosphere would be stimulated primarily by photons {with energies of $> 7.112$~keV}.
The fluorescence probability can be estimated from the photoabsorption cross section of atmospheric Fe at $\gtrsim 7.11$~keV ($< 4\times10^{-20}$~cm$^{2}$), column density ($\sim 2\times10^{8}$~cm$^{-2}$ by \citealt{yi09}), and the fluorescence yield ($\approx 30\%$ by \citealt{kaastra93}). The resultant probability is $< 10^{-10}$, meaning that the fluorescence line is too weak to explain the observations.
Thus, the fluorescence scenario is unlikely to explain the neutral/low-ionized Fe-K$\alpha$ lines.

\begin{figure}[htb!]
    \centering
    \includegraphics[width=8cm]{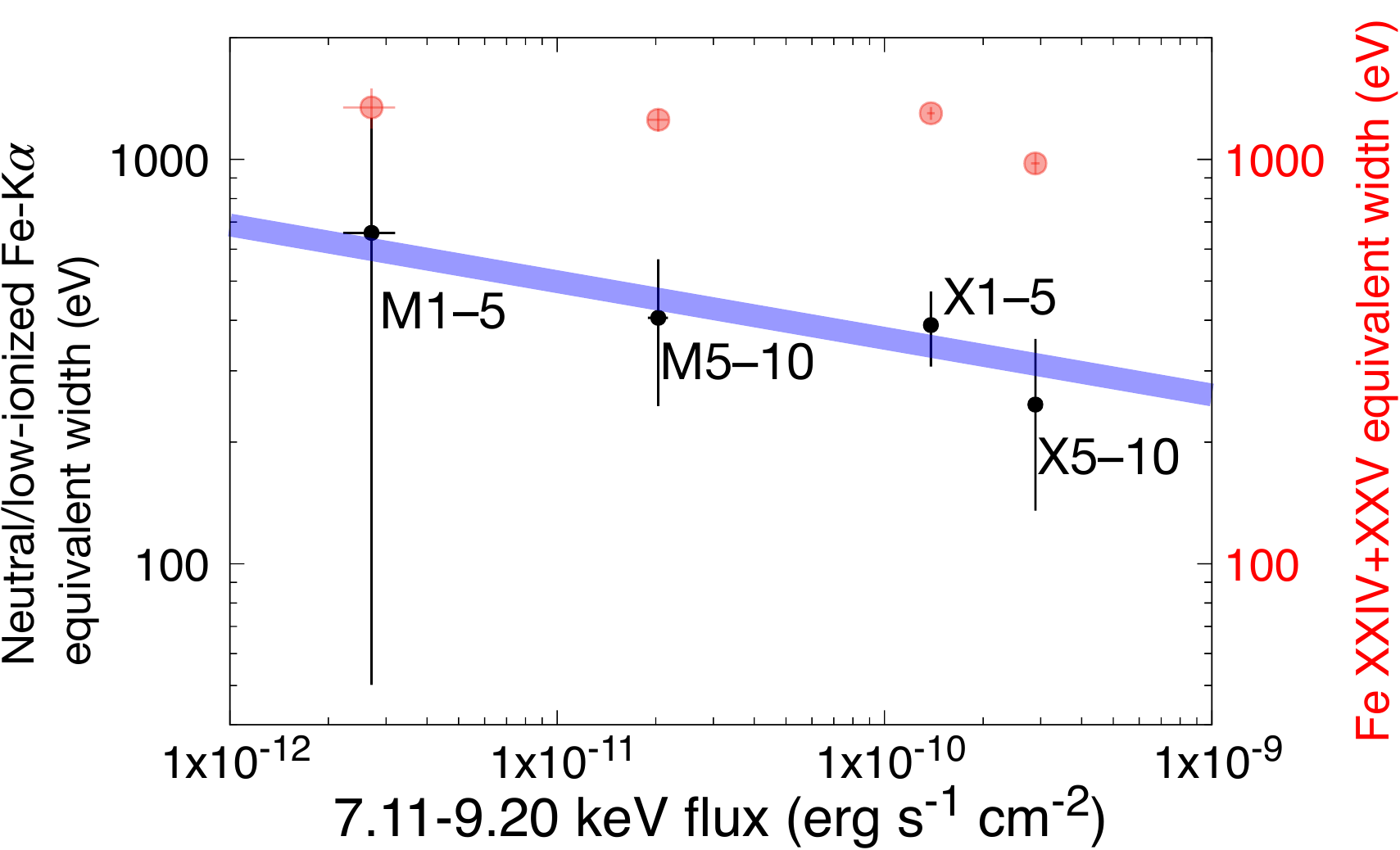}
    \flushleft
    \caption{Equivalent widths of the neutral/low-ionized Fe K$\alpha$ modeled by a Fe \emissiontype{XXI} line at $6.537$~keV (black) and Fe \emissiontype{XXIV}/\emissiontype{XXV} (red) obtained with Xtend as a function of 7.11--9.20~keV flux for four different flare classes. The blue line is the best-fit power-law function for the neutral/low-ionized Fe K$\alpha$ equivalent widths.
    Alt text: {Equivalent widths of two line emission over X-ray flux, shown for 10 to the minus 12 to 10 to the minus 9 erg per second per square centimeter.}}
    \label{fig-neutral-fe}
\end{figure}

If the neutral/low-ionized Fe-K$\alpha$ lines would originate from solar flares, further two possibilities can be considered as its origin: collisional ionization by non-thermal electrons accelerated in a flare loop \citep{osten10} or photoionization by hard X-ray photons emitted from a loop \citep{testa08}.
It will be meaningful to investigate the relation between the equivalent width of Fe-K$\alpha$ and flare magnitude, since an anti-correlation is expected in the photoionization scenario because larger loops would lead to weaker photon fluxes \citep{bai79, testa08, ercolano08}.
Figure~\ref{fig-neutral-fe} shows the Fe-K$\alpha$ equivalent width vs. 7.11--9.20~keV flux for four flare magnitudes.
The decreasing Fe-K$\alpha$ equivalent width as a function of flare magnitude, while the values of Fe \emissiontype{XXIV}+\emissiontype{XXV} are almost constant, suggests that the neutral/low-ionized Fe K$\alpha$ is stimulated by photoionization, although we cannot confidently rule out the electron collision scenario.
If we evaluate this trend with a power-law function, the power-law slope is determined to be $-0.14 \pm 0.09$.
This anti-correlation feature and the slope completely agree with a recent work on stellar flares ({$-0.27\pm0.10$ by \citealt{inoue25}}), implying that the same physical mechanism is operating in the Sun and magnetically active stars despite more than five order-of-magnitude discrepancy in X-ray flux.

{We note that the equivalent widths of fluorescence Fe K$\alpha$ may be somewhat higher than the typical expected values of $\sim 100$~eV, see, e.g., \cite{basko78, bai79}, where fluorescence emission from a cold matter with Solar abundances irradiated by isotropic continuum is calculated. An anisotropy of the irradiation (e.g., emission of non-thermal electrons can be beamed towards the Sun’s surface) or the inner-shell ionization by non-thermal electrons themselves (e.g., \citealt{tanaka84}) might explain this discrepancy. Also, Compton-scattered Fe \emissiontype{XXIV} + \emissiontype{XXV} lines on the Sun's surface may contribute significantly to the observed spectra. This component would result in a broad peak at energies below $\approx 6.6$~keV after reflection by the Earth atmosphere (see, e.g., \citealt{sunyaev96} for the discussion of Compton-scattered profiles in astrophysical conditions), which might explain the slight excess seen in Figure~\ref{fig-spec2}.
}



\subsection{Metal abundances of solar corona during flares}\label{sec-abund}
Our results on the metal abundance pattern are largely consistent with the preceding work with the same methodology \citep{katsuda20}, showing the lowest abundance of S and higher values of lower and higher FIP elements thus {partially} indicating the inverse-FIP effect {(Figure~\ref{fig-abund})}. The abundance pattern of X5--10 flares, {i.e., lower Fe, Si, S, and Ar, and higher Ca and Mg abundances than the other flare classes (Figure~\ref{fig-abund}),} is consistent with the results of \cite{katsuda20}, which used flare classes higher than X5.4.
On the other hand, the abundances of smaller flares are generally higher {except for Ca and Mg}.
The newly measured abundance of Fe was found to be similar to Si. These elemental abundances are in accordance with theoretical predictions \citep{laming04, laming12, laming15, laming21}, where low-FIP elements are enriched in the corona by the upward ponderomotive force.

This work has shown the abundance patterns for different flare classes spanning from M1 to X10. This dependence on flare class is characterized by the decreasing Si, S, and Ar abundances with increasing flare magnitude.
This trend is consistent with the abundance patterns of weaker flares (A and B classes) \citep{mondal21, nama23}, where the abundances of Mg, Si, and S during flares were found to be similar to the photospheric values.
{The behavior of the Si and S abundances are similar to that reported by \cite{woods23}, which showed a decreasing trend of the Mg, Si, S, Fe abundances with increasing flare temperature. The trends of the Mg and Fe abundances are different from our results. As mentioned in \cite{woods23}, systematics in treating multi-temperature plasma models can be a cause of this inconsistency.}
The theoretical model by Laming \citep{laming21} also agrees with these results, as stronger flares dredge up elements from deeper layers in the chromosphere, where higher depletion of {low-FIP} elements is expected.

The measured abundance of Ca, which is a factor of two to three higher than the photospheric value, is even higher than the theoretical prediction \citep{laming21}. Similar indication of high Ca abundances has been obtained in previous studies \citep{dennis15, katsuda20, sylwester84, sylwester98, sylwester22}.
In particular, the anti-correlation between the Ca abundance and flare magnitude over a wide range of C1- to X10-class flares \citep{sylwester22} is actually flat or possibly positive for the smaller range spanning from M1 to X10 classes and with a large scatter. Our measurements are also within this scatter range.
{As expected from the opposite behavior of the Ar and Ca abundances depending on flare class, the Ar/Ca ratio, which is an indicator of the FIP bias, decreases with increasing flare class. The value for the X5--10 class is consistent with those reported in \cite{katsuda20} and those for the lower classes are closer to the values seen in localized low-FIP depleted regions (inverse-FIP patches; \citealt{doschek15}).}
More detailed investigation of the flare-to-flare dependence will be important in understanding the behavior of the metal abundances. The mission lifetime of XRISM, designed to be three years but can be extended, is thus essential for such a systematic study of solar flares.

By tracing the short-term evolution of three X-class flares, we have found that the abundances of low-FIP elements, Si, S, Ca, and Fe, show an enhancement on a timescale of $\sim 100$~s ahead of flares (Figure~\ref{fig-abund2}).
This indicates that low-FIP elements are enriched at the beginning of flares, reflecting the coronal abundance pattern before it is mixed with chromospheric plasma with depleted low-FIP elements.
This trend is consistent with the findings by \cite{warren14}.
On a longer timescale of a few 1000~s after flare peaks, the Si and S abundances show an increasing trend as can be seen in Figure~\ref{fig-abund3}. Previous results on a similarly long-timescale behavior obtained by Suzaku \citep{katsuda20}, Chandrayaan-2 Solar X-ray Monitor \citep{mondal21, nama23}, Solar Dynamics Observatory \citep{warren14}, and {Miniature X-ray Solar Spectrometer \citep{woods23}} indicate that the abundances of low-FIP elements, Mg, Al, Si, and S are clearly depleted {relative to the photospheric values} during flares on a timescale of a few 1000~s. {This can be explained by the upflow of unfractionated chromospheric material.}
Then, the abundances increase during the decay phase of flares. This is believed to show recovering of the coronal abundances.
%
%
To summarize, our findings agree with theoretical and other observational studies. It is worth emphasizing that the large grasp of Xtend provides an excellent ability to trace the time evolution of the metal abundances.

\section{Conclusion}\label{sec-conclusion}
The astronomical satellite XRISM is observing solar flare X-rays reflected in the Earth's atmosphere during day-Earth occultation periods. Although day-Earth {occultation data} are a by-product of scientific observations, this work demonstrated that they are of great scientific importance themselves in solar physics.

We measured the abundances of Mg, Si, S, Ar, Ca, and Fe during M1--X10-class flares using a one-year data set by converting the observed equivalent widths of the line emission to the abundances considering the spectral deformation by reflection, confirming the inverse-FIP effect claimed by \cite{katsuda20}, which used the same methodology. We newly found a variation in the abundances of Si, S, and Ar with flare magnitude, which is consistent with the theoretical predictions by \cite{laming21}. The abundance of Ca shows an opposite trend, which may be mysterious but is consistent with previous work \citep{sylwester22}.
With the large grasp of Xtend, we successfully traced a time variation of metal abundances with a 100-s resolution, providing evidence for enriched low-FIP elements before flare peaks.

The Fe-K emission lines were successfully decomposed into Rayleigh- and Compton-scattered Fe \emissiontype{XXIV}/\emissiontype{XXV} lines and neutral/low-ionized Fe lines. The Rayleigh-to-Compton ratio was roughly consistent with the average scattering angle of the observation data. The equivalent width of the neutral/low-ionized Fe K$\alpha$, which was measured for the first time since the era of Hinotori, was found to show an anti-correlation with flare magnitude and 7.11--9.20~keV flux. This supports the idea that Fe-K$\alpha$ is stimulated by hard X-ray photons emitted from flare loops.

This work demonstrated that XRISM can be a powerful tool in the field of solar physics, offering valuable high-statistic CCD data and high-resolution microcalorimeter spectra in the energy range covering the Fe-K band.

\section*{Acknowledgement}
{We appreciate the helpful suggestions by the referee, which have improved the quality of this paper significantly.}
This work was partly supported by JSPS KAKENHI grant numbers 24K17093, 21H01095, 23K20850, and 24H00253, and JSPS Core-to-Core Program (grant number:JPJSCCA20220002). IK acknowledges support by the COMPLEX project from the European Research Council (ERC) under the European Union’s Horizon 2020 research and innovation program grant agreement ERC-2019-AdG 882679.

\bibliography{references.bib}
\bibliographystyle{test_pasj}

\appendix
\section{Pre-flare Background Spectra}\label{sec-bgd}
\begin{figure}[htb!]
    \centering
    \includegraphics[width=8cm]{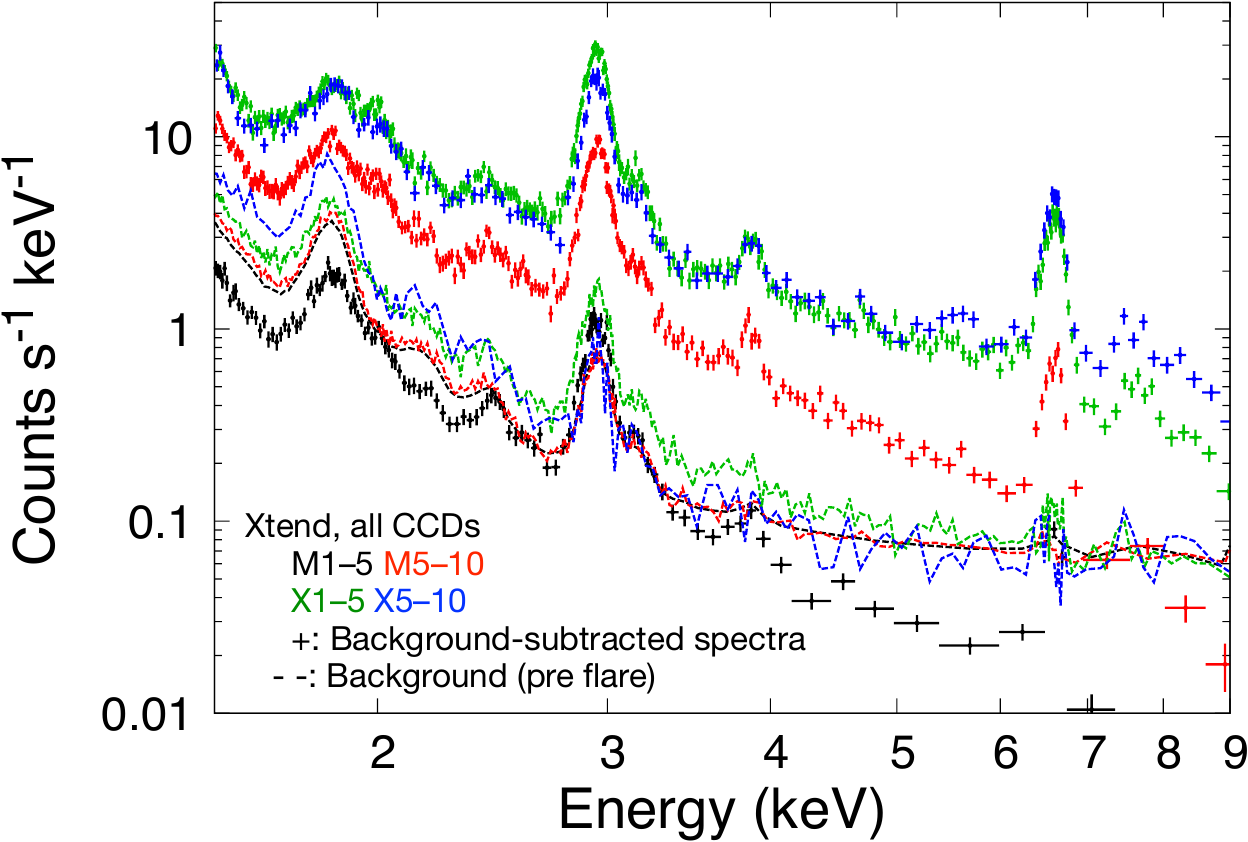}
    \flushleft
    \caption{Pre-flare background spectra (dashed lines) and background-subtracted flare spectra (crosses) for four different flare magnitudes.
    Alt text: {Four flare spectra and corresponding background spectra. The background level is higher than the flare spectrum only in the M1 to 5 class flares.}}
    \label{fig-bgd}
\end{figure}

{Figure~\ref{fig-bgd} shows the pre-flare background spectra we used in our abundance measurement in Section~\ref{sec-ana-abund} in comparison with the background-subtracted flare spectra.}

\section{Correlation between the $\alpha$ parameter of the {power-law} DEM model and X-ray Flare Class}\label{sec-alpha}
\begin{figure}[htb!]
	\centering
	\includegraphics[width=8cm]{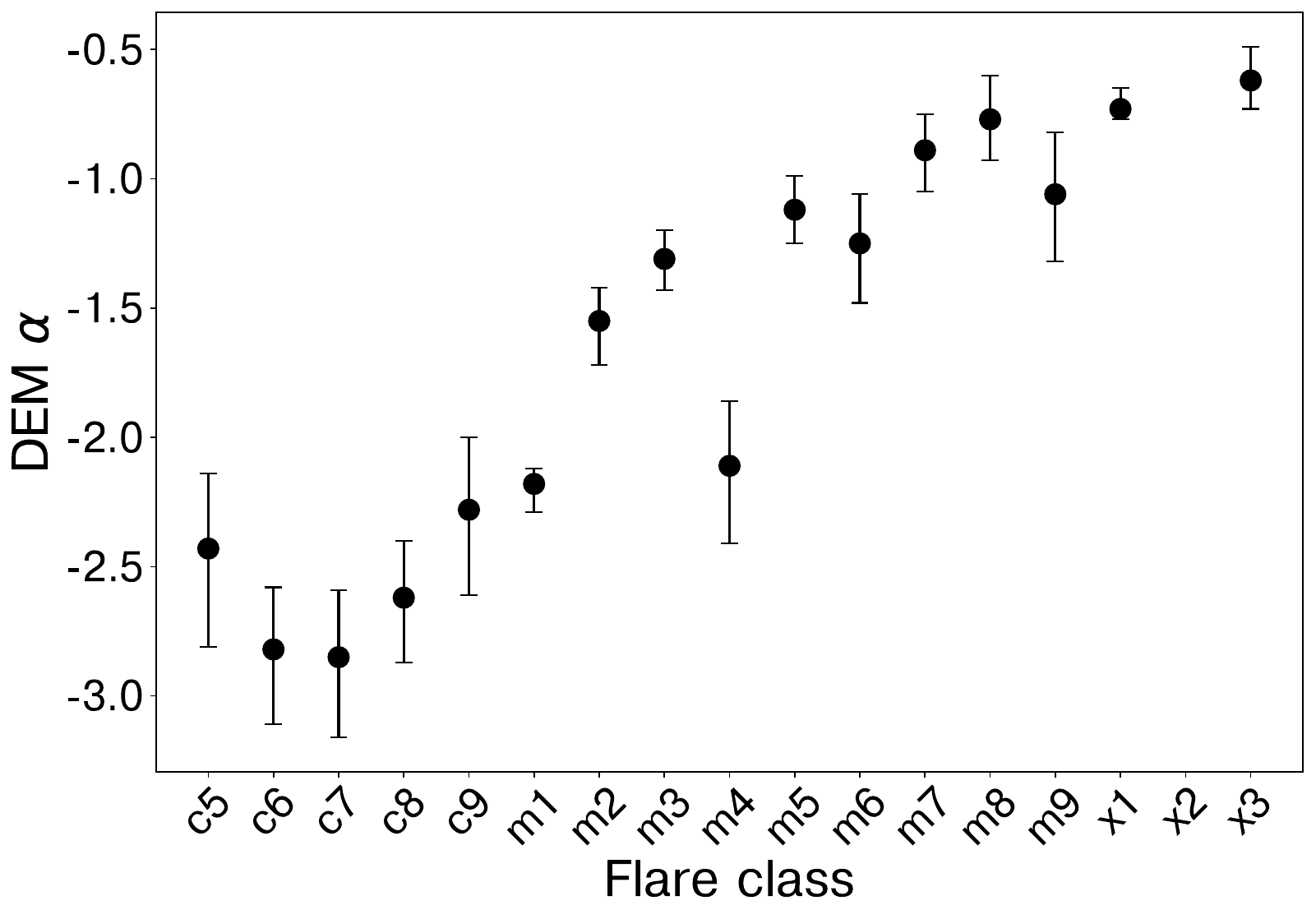}
	\flushleft
	\caption{The $\alpha$ parameter of the {power-law} DEM model {(DEM $\alpha$)}, which is an indicator of the flare temperature, shown as a function of flare class.
    Alt text: {The alpha parameter against flare class, which shows a positive correlation.}}
	\label{fig-alpha}
\end{figure}

In addition to the abundance pattern and its dependence on the flare class discussed in Section~\ref{sec-abund}, this appendix presents the correlation between the $\alpha$ parameter of the {power-law} DEM model (DEM $\alpha$) and flare class. Figure~\ref{fig-alpha} shows DEM $\alpha$ as a function of flare class. Here we use day-Earth {occultation data} with Xtend from October 2023 to December 2024, covering 421 flares ranging from C5 to X3. To use data near the flare peaks, we select time intervals of $\pm$2000 seconds around the peak time of the GOES X-ray flux. The calculation of DEM $\alpha$ follows the method described in Section~\ref{sec-ana-abund}, based on \cite{katsuda20}. We find that DEM $\alpha$ is positively correlated with flare class, suggesting that higher flare classes are associated with a stronger contribution from high-temperature components. 

\begin{figure}[htb!]
	\centering
	\includegraphics[width=8cm]{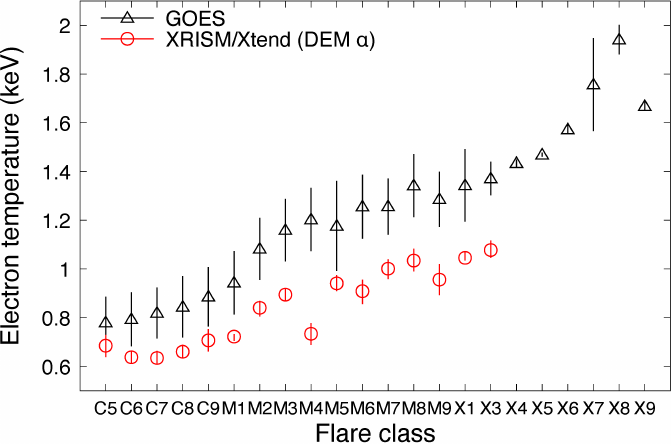}
	\flushleft
	\caption{{Electron temperatures derived with GOES and Xtend (based on DEM $\alpha$) as a function of flare class. The temperature corresponding to DEM $\alpha$ is the representative temperature of the collisional ionization equilibrium plasma model ({\tt apec}) that best reproduces the DEM spectrum at the Si \emissiontype{XIII} and \emissiontype{XIV} lines. 
    Alt text: Two temperature estimates versus flare class, both of which show an increasing trend.}}
	\label{fig-alpha2}
\end{figure}

{We compare the temperatures corresponding to the DEM $\alpha$ values with the temperatures measured based on the GOES data. The conversion from DEM $\alpha$ to temperature is calculated by finding the temperature of the collisional ionization equilibrium plasma model ({\tt apec}) that best reproduces the DEM spectrum at the Si \emissiontype{XIII} and \emissiontype{XIV} lines. The temperatures from the GOES data are calculated for the periods of $\pm30$~seconds of the flares.
Figure~\ref{fig-alpha2} shows the results. The two temperatures show the same increasing trend with an increasing flare magnitude with an offset of $\sim 0.2$~keV. This is reasonable given that GOES data are sensitive to higher temperature components than Xtend due to the different sensitive energy ranges.
The temperatures based on the Xtend data of the X4 and higher-class flares are not included because of the poor statistics or lack of the data.
It is worth comparing the temperature of 1.5~keV, which is assumed in our Fe-K spectral modeling, with the average temperature of the X-class flares, which is found to be $1.4\pm0.2$~keV (the error indicates the standard deviation) based on the GOES data. We note that the temperature determined with GOES better approximates the representative temperature in the Fe-K energy range than the Xtend data.}

\end{document}